\documentclass[aps,pre,twocolumn,floatfix]{revtex4-2}
\usepackage[english]{babel}
\usepackage{float}
\usepackage{amsmath}
\usepackage{comment}
\usepackage{kantlipsum} 
\usepackage{graphicx}
\usepackage{xcolor}
\usepackage[normalem]{ulem}
\usepackage{color}
\definecolor{darkgreen}{rgb}{0.0, 0.4, 0.0}
\definecolor{oooo}{rgb}{0.99, 0.5, 0.01}

\begin{document}

\title{Elongated particles discharged with a conveyor belt in a two-dimensional silo}

\author{Bo Fan\textit{$^{1,3}$}}
\author{Iker Zuriguel\textit{$^{2}$}}
\email{iker@unav.es}
\author{Joshua A. Dijksman\textit{$^{3,4}$}}
\author{Jasper van der Gucht\textit{$^{3}$}}
\author{Tam\'as B\"orzs\"onyi\textit{$^{1}$}}

\address{
$^1$Institute for Solid State Physics and Optics, Wigner Research Centre for
Physics, P.O. Box 49, H-1525 Budapest, Hungary\\
$^2$F\'isica y Matem\'atica Aplicada, Facultad de  Ciencias, Universidad de Navarra, Pamplona, Spain\\
$^3$Physical Chemistry and Soft Matter,  Wageningen University $\&$ Research, Wageningen, The Netherlands \\
$^4$Van der Waals-Zeeman Institute, Institute of Physics, University of Amsterdam, Science Park 904, 1098 XH Amsterdam, The Netherlands\\
}

\vspace{10pt}

\begin{abstract}
The flow of elliptical particles out of a 2-dimensional silo when extracted with a conveyor belt is
analyzed experimentally. 
The conveyor belt - placed directly below the silo outlet - reduces the flow rate, increases the size of the stagnant zone, and it has a very strong influence on the relative velocity fluctuations as they strongly increase everywhere in the silo with decreasing belt speed. In other words, instead of slower but smooth flow, flow reduction by belt leads to intermittent flow.
Interestingly, we show that this intermittency correlates with a strong reduction of the orientational order of the particles at the orifice region. Moreover, we observe that the average orientation of the grains passing through the outlet is modified when they are extracted with the belt, a feature that becomes more evident for large orifices.  
\end{abstract}

\maketitle

\section{Introduction}

Granular flows are ubiquitous in nature, and frequently, display unpredictable phenomena. Indeed, we all have observed uncontrolled clogging behavior and intermittent flow when pouring a granulate eg. salt, sugar or cereals. Flow of a granulate out of a container or a silo is widely used in various industrial applications. A  very useful feature of such flows is that the flow rate does not depend on the filling height of the silo, as it was already pointed out in pioneering works long ago~\cite{Hagen,Tighe}. For large enough orifices there is a clear relation between the size of the orifice and the flow rate~\cite{Hagen,beverloCES1961,mankocGM2007}. Thus, one can tune the flow rate by simply changing the orifice size. Unfortunately, this only works for relatively large flow rates, because below a certain orifice size the system clogs \cite{zuriguelPRE2005,thomasPRE2013,thomasPRL2015}. In industrial applications, however, one often needs small flow rates. There are several ways to overcome this problem, e.g. reducing the probability of clogging by vibration \cite{martinezPRE2008,mankocPRE2009,lozanoPRL2012,zuriguelSR2014}, by placing an obstacle above the orifice to prevent arch formation \cite{zuriguelPRL2011,endoPRF2017,reddyPRE2018}, or simply using a conveyor system (screw conveyor, or conveyor belt) placed directly below the silo outlet so that it limits the flow rate \cite{fernandezCES2011}. Then, one can use a large orifice (clogging is avoided) and the flow rate is set by the speed of the conveyor system.

In this work we investigate how the presence of a conveyor belt influences the flow field, packing fraction, and grain orientation during the discharge of a 2-dimensional (2D) silo filled with ellipses. In previous works~\cite{gellaPRL2018,Gella2019PRL,gellaPT2020}, for the case of spheres, it was proved that the presence of a belt strongly affects these physical fields. In particular, a transition was shown from belt-controlled flow to outlet-size controlled flow (by gravity) as the belt velocity increased and the orifice reduced. The advantage of 2D model silos is that one can obtain detailed microscopic information about the flow and clogging of the granulate by using high speed digital imaging \cite{rubiolargoPRL2015}. Such access to the internal flow structure is more  difficult to obtain in 3-dimensional systems,  where one has to use more sophisticated detection techniques (e.g. X-ray tomography)~\cite{borzsonyiNJP2016,guillardSR2017,stannariusNJP2019}. 

The majority of previous experimental or numerical works focusing on microscopic details of silo discharge used spherical (3D) or disk shaped (2D) particles. Some studies provided data for shape-anisotropic grains~\cite{clearyAMM2002,taoCEP2010,liuPT2014,borzsonyiNJP2016,guillardSR2017,szaboPRE2018}.
For elongated particles, orientational ordering develops in the sheared regions \cite{borzsonyiNJP2016}, where the average alignment of the particles' long axis is nearly parallel to the flow direction, similarly to the observations in other shear flows \cite{borzsonyiPRL2012,borzsonyiPRE2012}. For such grains, the flow field in the silo showed stronger funneling and larger velocity fluctuations \cite{clearyAMM2002,taoCEP2010,liuPT2014,borzsonyiNJP2016,guillardSR2017,szaboPRE2018} and the probability of clog formation was also higher \cite{ashourSM2017,reddyJSM2021} than for spherical grains. Therefore, as a natural extension of these works, in this study we focus on determining the influence of the conveyor belt extraction system on the flow of elliptical grains, with special focus on grain orientation, ordering and the flow fields. 

\section{Experimental setup}
\begin{figure}[htbp!]
    \centering
    \includegraphics[width=1.0\linewidth]{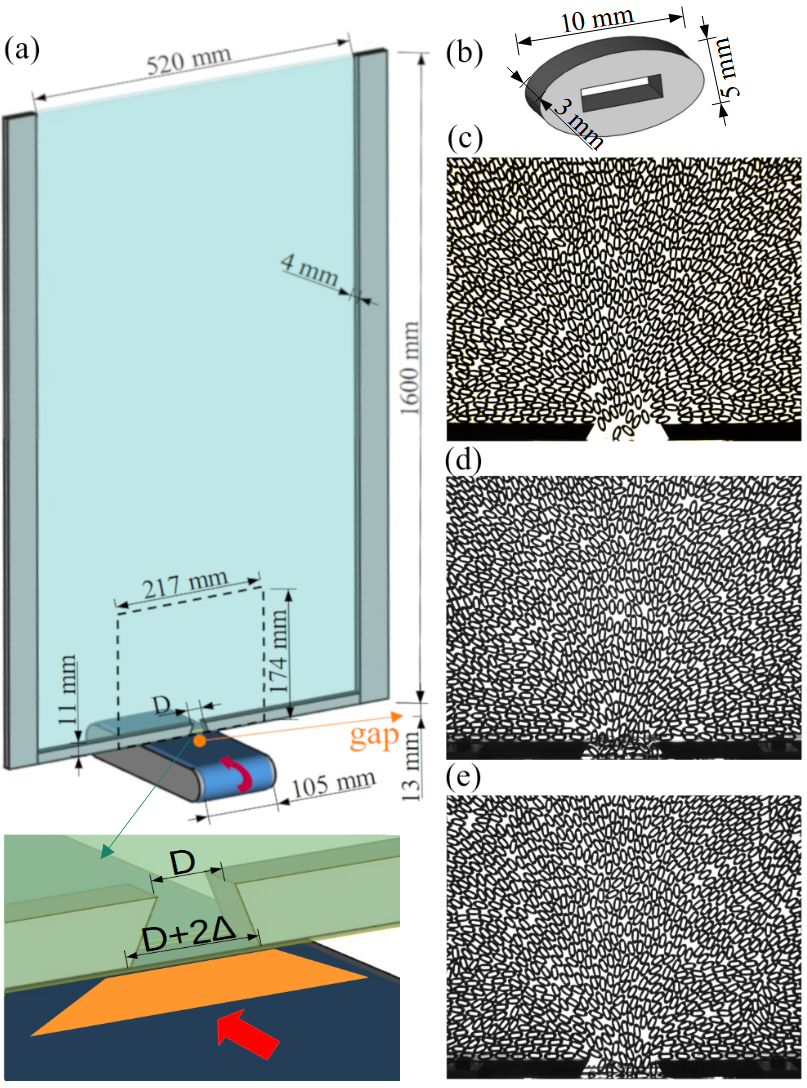}
    \caption{Experimental setup. (a) A sketch of the experimental setup. The rectangle area formed by the dashed lines is the camera's scope. For belt controlled silo discharge the flow expands below the orifice due to the trapezoid shape of the outlet characterized by the parameter $\Delta=5.7$ mm and the non-zero gap ($13$ mm) between the bottom of the silo walls and the belt. On the right column, (b) sketch of the elliptical particle, photos of (c) the free flow discharge (without the belt below the outlet), and flow-rate-controlled flow with belt speeds of (d) $5.1$ mm/s and (e) $55.7$ mm/s. The orifice size in the photos is the same, which is $40$ mm.}
    \label{fig:setup}
\end{figure}

We use the same experimental setup (Fig.~\ref{fig:setup}~(a)) and protocol implemented in previous works using spherical particles \cite{gellaPRL2018,Gella2019PRL,gellaPT2020}. 
The system comprises a two-dimensional (2D) silo, a removable conveyor belt below the silo outlet and a high-speed camera focusing on the lower part of the silo above the orifice (the rectangle area drawn with dashed lines). The silo is formed by two parallel glass plates with a height of $1600$ mm, a width of $700$ mm and a thickness of $6$ mm, between which there are two vertical $4$ mm thick and $40$ mm wide aluminum bars at both sides with a distance of $520$ mm. The particles filled 
in the silo are elliptical cylinders (see Fig.~\ref{fig:setup}(b)) of 10 mm in length, 5 mm in width, and 3 mm in thickness. Therefore, a single layer can fit in the 4 mm gap between the two glass plates.  
In addition, the width $D$ of the silo outlet at the bottom can be adjusted with two $11$ mm wide horizontal sliders. Lastly, a conveyor belt with a width of $105$ mm is mounted below the silo outlet with a gap of approximately $13$ mm between the glass plate edge and the conveyor's top surface to control the outflow rate. For comparison, we have also implemented experiments in which the conveyor belt is not mounted, hence obtaining the so called free-flow case. 
In Fig.~\ref{fig:setup}~(c)-(e), photos from the experiments with a $40$ mm orifice are presented for free-flow and flow-rate-controlled flow with a conveyor belt speed at $5.1$ mm/s and $55.7$ mm/s. 
We note that when using identical spheres hexagonal ordering of the grains was observed (see Fig. 1(b) in Ref. \cite{gellaPT2020}), which is not present here for the case of elliptic particles.
In this work, four belt speeds were used: $1.5$ mm/s, $5.1$ mm/s, $10.5$ mm/s, $55.7$ mm/s. Note that 1.5 mm/s corresponds to the so-called quasi-static regime in Refs.~\cite{gellaPRL2018,Gella2019PRL,gellaPT2020}.

The silo was filled from the top with elliptical particles up to a height of about $40$ cm. We used a high-speed camera at a frequency of either $1000$ Hz (for free flow) or $125$ Hz (for flow-rate-controlled flow) to capture the movement of grains in a rectangular area (marked with a dashed-line in Fig.~\ref{fig:setup}~(a)) while the grains are discharged through an outlet of $D=40$ mm, $50$ mm, $60$ mm or $80$ mm. The data presented in this work are based on 3 discharges for free flow and 1 discharge for each belt velocity for the case of belt controlled flow, and the number of images N analyzed for each configuration is in the range of $5000 < N < 36000$, with a larger number of images for the belt controlled cases. The flow field, grain orientations and packing fraction were determined by digital image analysis, and the flow rate was measured by counting the number of grains passing through the orifice line.

\section{Results and discussion}
\begin{figure*}[htbp!]
    \centering
    \includegraphics[width=1.0\textwidth]{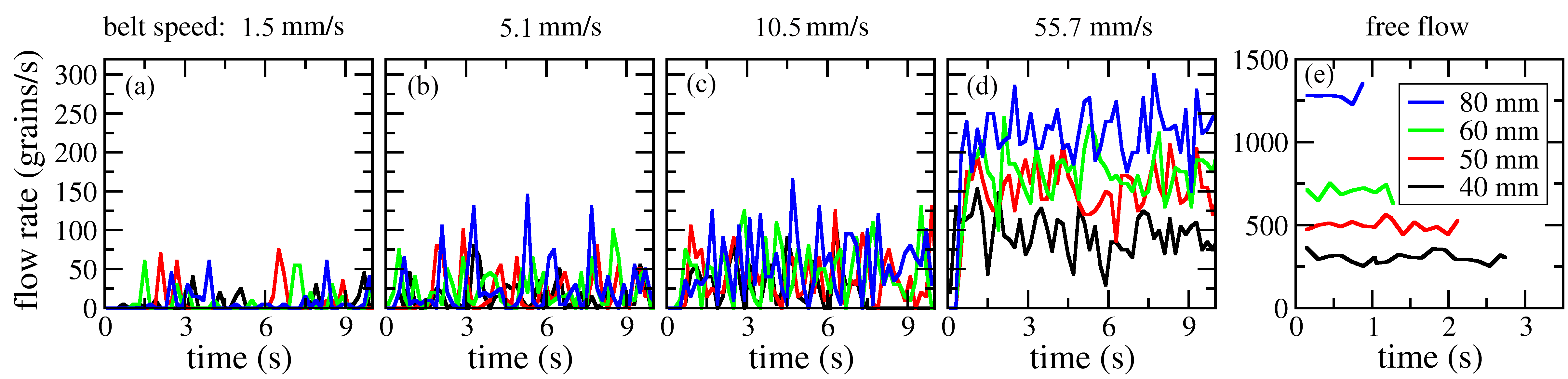}
    \caption{Flow rate evolution for different orifice sizes ($40$ mm, $50$ mm, $60$ mm and $80$ mm) as indicated in the legend of panel (e). (a)-(d) Flow with a conveyor belt speed of $55.7$ mm/s, $10.5$ mm/s, $5.1$ mm/s, and $1.5$ mm/s respectively, (e) free flow.   
    }
    \label{fig:flow_rate}
\end{figure*}

\begin{figure*}[htbp!]
    \centering
    \includegraphics[width=0.85\linewidth]{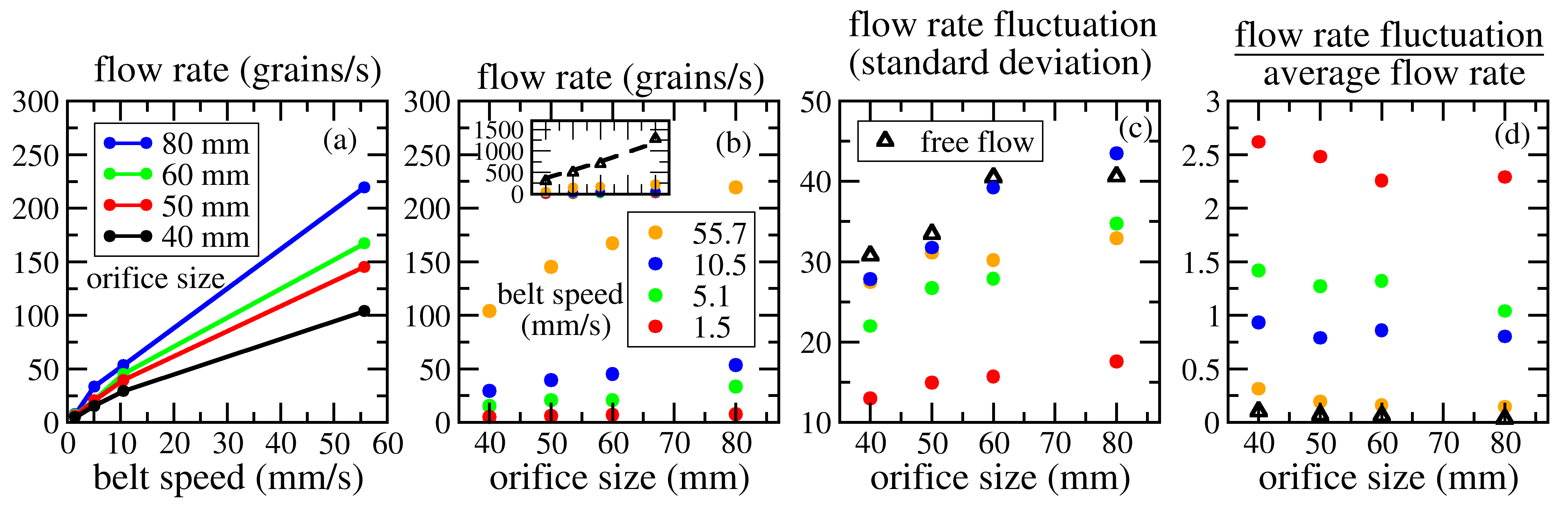}
    \caption{(a) Time averaged flow rate as a function of belt speed (different colors correspond to different outlet sizes). (b) Time averaged flow rate as a function of orifice size. In the inset the flow rate of the free-flow case (open triangular symbols) is compared with the flow-rate-controlled cases (colored solid circular symbols). (c) Flow rate fluctuation characterized by the standard deviation of the flow rate data shown in Fig.~\ref{fig:flow_rate}. (d) Relative flow rate fluctuation defined as the ratio of the standard deviation to the average flow rate.
    }
    \label{fig:flow_rate-orifice-belt}
\end{figure*}

The flow rate is shown as a function of time for the
case of free flow in Fig.~\ref{fig:flow_rate}(e) and for flow-rate-controlled
cases with different belt speeds in Figs.~\ref{fig:flow_rate}(a-d). Clearly, the flow rate is considerably reduced by the presence of the belt while the fluctuations seem to decrease only moderately.
In order to better quantify this, we represent the average flow rate as a function of belt speed (Fig.~\ref{fig:flow_rate-orifice-belt}(a)) and the orifice size (Fig.~\ref{fig:flow_rate-orifice-belt}(b)). The results obtained reveal the same main features reported for the case of spheres (see Fig. 2 of Ref.~\cite{gellaPT2020}); 
i.e. i) the flow rate grows with the belt speed in a non-linear manner (Fig.~\ref{fig:flow_rate-orifice-belt}(a)) with deviations from linearity being more important as the belt speed ($v_b$) increases; ii) the flow rate is strongly dependent on the belt velocity, and even for the fastest case, is far from approaching the values of the free flow (see inset of Fig.~\ref{fig:flow_rate-orifice-belt}(b)). Concerning the fluctuations, their characterization by means of the standard deviation (Fig.~\ref{fig:flow_rate-orifice-belt}(c)) evidences an increase with the outlet size and the belt speed. However, when normalized by the average value of the flow rate (Fig.~\ref{fig:flow_rate-orifice-belt}(d)), the tendency is inverted, and the rescaled value reduces with both the outlet size and the belt  speed. As expected, when the grains are discharged in a quasi-static manner ($v_b=1.5$~mm/s), the fluctuations become much larger than the average.
\begin{figure*}[htbp!]
    \centering
    \includegraphics[width=0.93\textwidth]{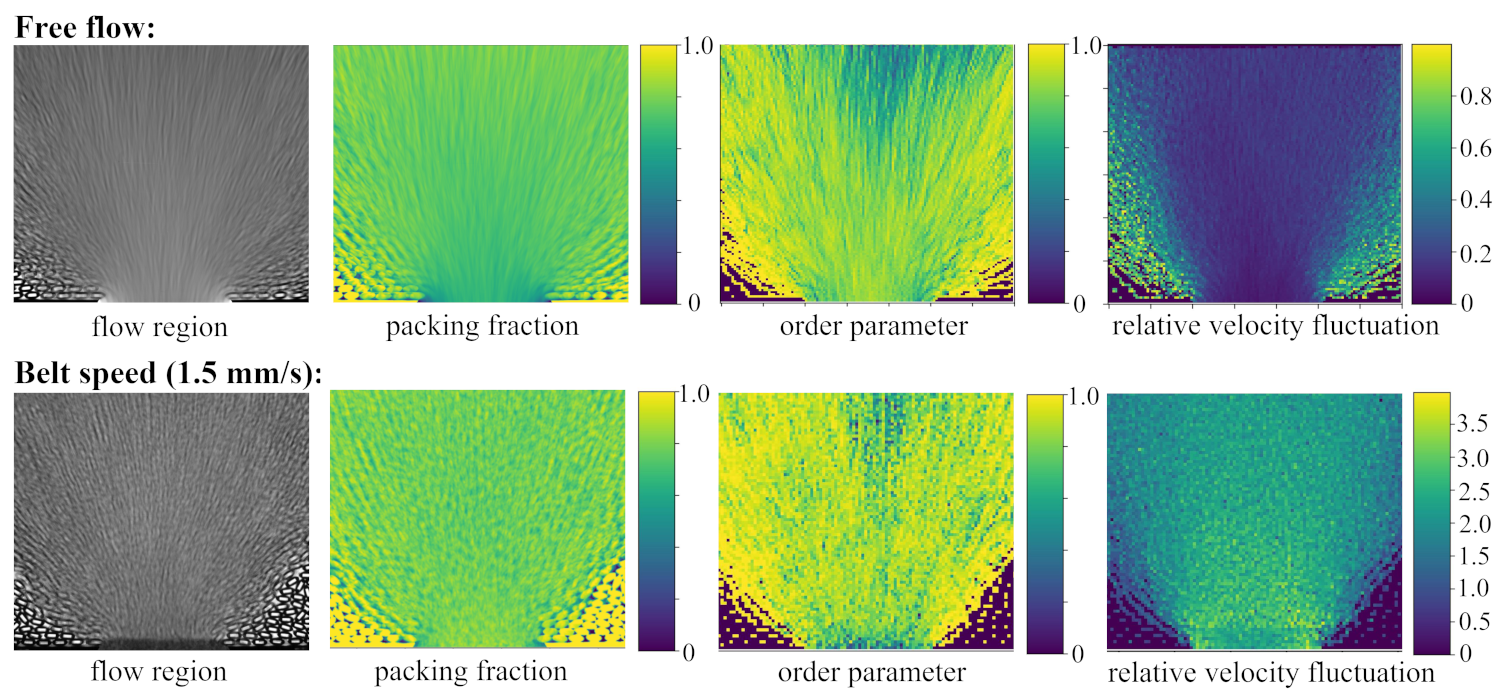}
    \caption{Flow field maps from left to right: flow region (superpositioned image to show the flowing particles), packing fraction, order parameter, and relative velocity fluctuation. The relative velocity fluctuation is defined as the ratio of the velocity standard deviation to the average velocity. The orifice size is always $80$ mm in these maps.}
    \label{fig:flow_field_maps}
\end{figure*}

After characterizing the general features of  the flow rate, we now turn to investigate the spatial structure of the flow field. In Fig.~\ref{fig:flow_field_maps} we present the flowing region (by overlapped images), the packing fraction, order parameter and the relative velocity fluctuations for free flow (top row) and for the quasi-static discharge ($v_b=1.5$~mm/s) in the rectangular area shown in Fig.~\ref{fig:setup}.
Orientational ordering is measured by the usual nematic order parameter S, which is defined as the average of $2\cos^2(\theta-\Bar{\theta})-1$ in a certain area, where $\theta$ is the orientation of individual particles and $\Bar{\theta}$ is the average orientation of the particles in the selected area.

As we see in the left column of Fig.~\ref{fig:flow_field_maps}, the stagnant zone on the two sides of the flowing region is considerably larger for the case of flow controlled by the conveyor belt than for free flow. Note that for correct comparison we constructed these 2 images by overlapping the snapshots obtained within a given time period in such a way that the number of grains leaving the silo was very similar (about 2300). For free flow this corresponds to 1765 images and a flowing time of 1.77 s. For the case of flow with belt, those snapshots where nothing moved in the silo were not included, thus the overlapped image corresponds to 5540 images and a flowing time of 44.32 s. 
The plots for the packing fraction and orientational order parameter (second and third columns of Fig.~\ref{fig:flow_field_maps}) both show that the presence of the conveyor belt influences these quantities mostly in the orifice region. 
For free flow there is a region above the orifice in which the grains accelerate towards the orifice and consequently the packing fraction is smaller compared to other regions in the silo (greener
 in the map). 
On the contrary, when we reduce the flow rate with the conveyor belt, the packing fraction in this region becomes large (yellower), reaching a value similar to what we observe elsewhere in the silo. Looking at the maps of the orientational order parameter, we see that in the outlet region (where the belt makes the packing denser) the order is decreased. As we will see later on, the shear induced orientational order (the grains above the orifice are typically aligned with their long axis towards the orifice) is destroyed by the belt, since the orientation of the grains changes in the orifice. 

Finally, we investigate the relative velocity fluctuations. This quantity is defined as the standard deviation of the flow velocity divided by the average flow velocity and is presented in the last column of Fig.~\ref{fig:flow_field_maps}. Clearly, the relative velocity fluctuations get significantly increased when the grains are discharged with the belt, a phenomenon that occurs in the majority of the flowing region and not only at the orifice. This means that reducing the flow speed by the belt results in slower but more fluctuating flow all throughout the silo. 

Despite this, since the influence of the conveyor belt on the other variables (such as packing fraction and order parameter) is mostly localized in the orifice region, we proceed to its characterization in the following. We study the 
profiles at the orifice of velocity, packing fraction, particle orientation and order parameter, first for the free-flow case, and then for the discharge with a conveyor belt.
\begin{figure*}[htbp!]
    \centering
    \includegraphics[width=1.0\textwidth]{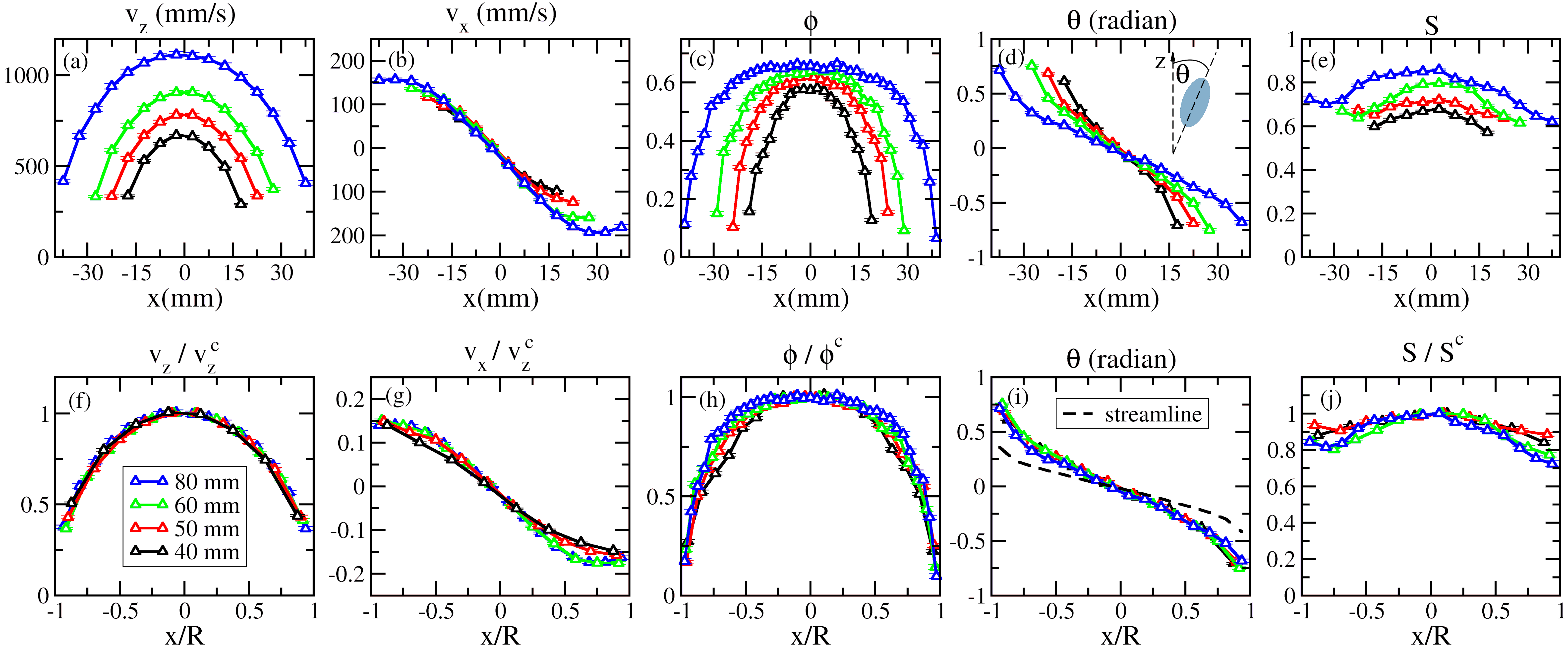}
    \caption{Profiles of different quantities at the orifice for the free flow case. (a)-(e) original data and (f)-(j) normalized data for vertical velocity $v_z$, horizontal velocity $v_x$, packing fraction $\phi$, average orientation $\theta$ and order parameter $S$, respectively. $x$ denoted on the x axis represents the position at the outlet line. Radius $R$ is the half of the corresponding orifice size $D$. $v^c_z$, $\phi^c$ and $S^c$ are the values of vertical velocity, packing fraction and order parameter at the orifice center respectively. The orientation of the particle is defined as the angle between the major axis of the particle and the vertical direction, as it is illustrated by the sketch in panel (d). The dashed line in (i) shows the direction of the streamline at the orifice as a reference.  Order parameter in (e) and (j) is a quantity measuring the extent of alignment of the particles, defined in the range between $0$ and $1$, corresponding to random orientation and perfect alignment, respectively. Different colors correspond to different orifice size as indicated in the legend of panel (f).}
    \label{fig:free_flow}
\end{figure*}
\begin{figure}[htbp!]
    \centering
    \includegraphics[width=0.8\linewidth]{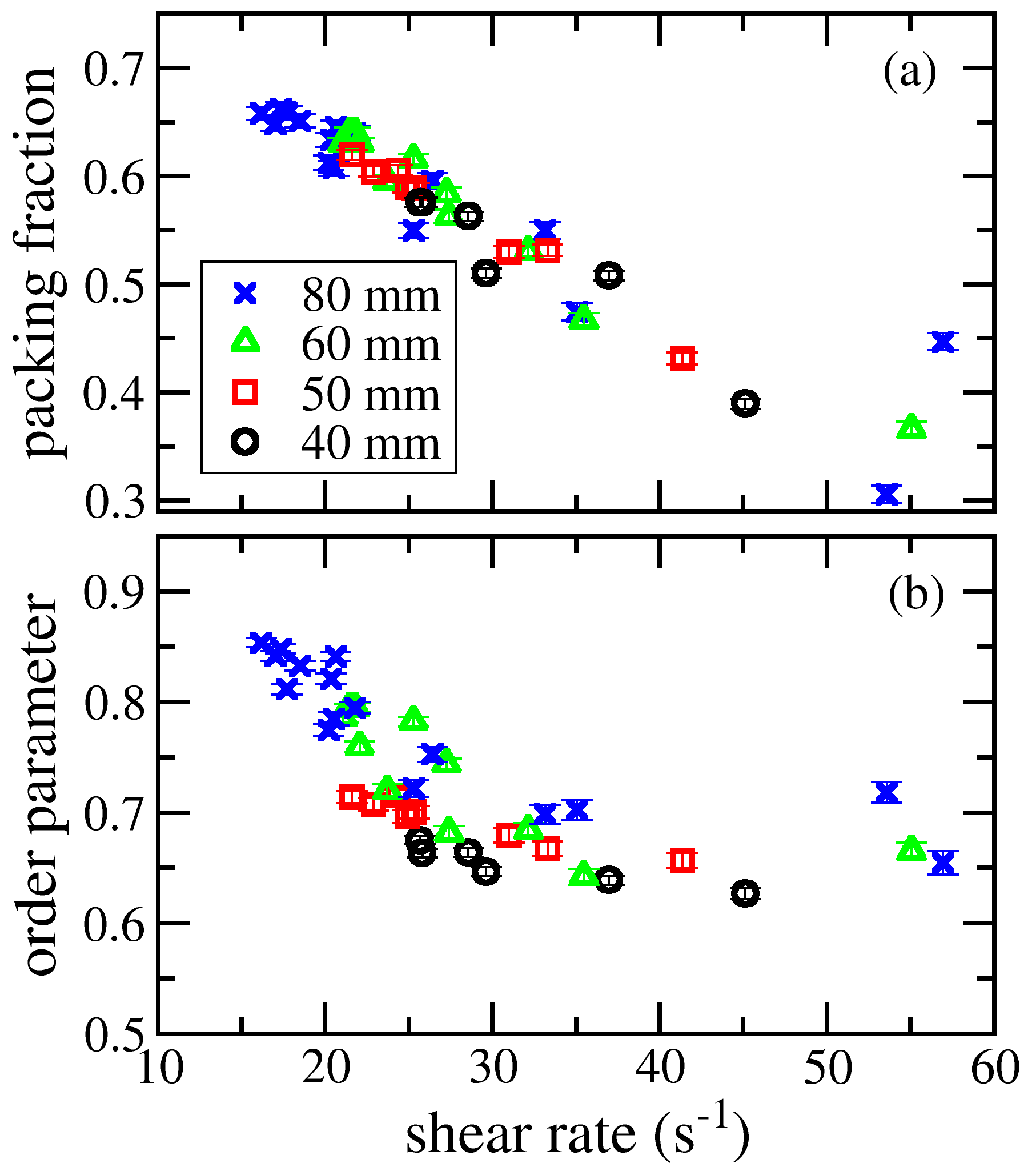}
    \caption{Free flow: (a) packing fraction, and (b) order parameter as a function of shear rate, which is defined as the absolute value of the velocity gradient of the flow field.}
    \label{shear_rate-phi-s}
\end{figure}
\subsection{Free flow}
\label{free-flow}

In Fig.~\ref{fig:free_flow}, we evidence that the vertical velocity (Fig.~\ref{fig:free_flow}(a) and (f)) and the packing fraction profiles (Fig.~\ref{fig:free_flow}(c) and (h)) of the elliptical particles show similar trends to those reported for spherical particles in Ref.~\cite{janda2012PRL}. In short, the vertical velocity (Fig.~\ref{fig:free_flow}(a)) gets higher as the orifice size is increased, and the profiles of vertical velocity $v_z$ have a universal shape (Fig.~\ref{fig:free_flow}(f)), since the data nicely collapse when the horizontal coordinate is divided by the half size of the orifice ($R$) and the vertical velocity is normalized by its maximum value at the center of the silo ($v^z_c$). These collapsed datasets agree with the function of $v_z/v^c_z = \sqrt{1-(x/R)^2}$ which originates from the assumption of the existence of a parabolic free fall arch above the orifice~\cite{janda2012PRL}. 

The packing fraction (Fig.~\ref{fig:free_flow}(c)) shows a similar trend to the vertical velocity: the packing is denser in the middle of the orifice than near the edge. This suggests a correlation between the shear rate and the packing, which is not surprising as larger shear rate means more intensive collisions and thus the expansion of the material. In Fig. \ref{shear_rate-phi-s}(a) we evidence this feature by representing the packing fraction as a function of shear rate, which is defined as the absolute value of the gradient of the velocity field. We see a reasonable overlap of the curves obtained for different orifice sizes. 
Again, non-dimensionalizing the packing fraction curves shown in Fig.~\ref{fig:free_flow}(c) as we did with the velocity profiles, the data approximately collapse to a single curve (see Fig.~\ref{fig:free_flow}(h)), although not as precisely as the vertical velocity case.
We see slightly smaller packing fractions when the orifice size reduces.

Regarding the horizontal velocity, in Fig.~\ref{fig:free_flow}(b) we observe that the particles move towards the orifice center. It is noteworthy that the curves measured for different orifice sizes 
fall very close to each other, with the horizontal velocity increasing with the distance from the orifice center. Normalizing the horizontal velocity by the maximum of the vertical velocity also leads to a reasonably good collapse, and helps to visualize that the maximum horizontal velocity (at the edge of the orifice) is about 10 \% of the maximum vertical velocity.

For a complete description of the flow of elliptical particles, in addition to their position, velocity and packing fraction, we also need to characterize their orientation, which is measured by the angle between the major axis of the ellipse and the vertical direction.
Fig.~\ref{fig:free_flow}(d) and (i) show the distribution of particle orientation at the outlet line in the original and normalized forms respectively. Once more, when the $x$ axis is normalized by $R$, the curves of the average orientation at the outlet line overlap. Remarkably, the collapsed curves are near to the streamline direction curve,
(dashed line in Fig.~\ref{fig:free_flow}(i), which is obtained from the average velocity direction). 
A sketch in Fig.~\ref{fig:dgm_orientation}(a) visualizes the typical average grain orientations. This is in accordance with the observations on the direction of the elongated particles under shear in 3D or in confined systems, where the long axis of the particle in average was found to be aligned near to the flow direction ~\cite{borzsonyiPRL2012,borzsonyiPRE2012,borzsonyiSM2013,polNJP2022}. In the silo, this means that the orientation of the grains with respect to vertical is slightly more tilted than the flow lines  themselves \cite{borzsonyiNJP2016}.

The order parameter (Fig.~\ref{fig:free_flow}~(e)) is larger in the middle of the orifice than near the orifice edge. Similarly to the packing fraction, the order parameter correlates with the shear rate: larger shear rate leads to decreased order parameter, as we see in Fig. \ref{shear_rate-phi-s}(b). This tendency is also consistent
with previous observations in shear flows ~\cite{borzsonyiPRL2012,borzsonyiPRE2012}. Thus, more intensive collisions lead to expansion of the material and decreased orientational ordering. Again, Fig.~\ref{fig:free_flow}(j) shows that similarly to what occurs with the vertical velocity, a rescaling of the horizontal coordinate by
R and the order parameter by its maximum value measured in the center
of the silo ($S^c$), leads to a reasonably good collapse of the data. 

In summary, the results of this section corroborate that, as it occurs for spheres, in the free discharge of elliptical particles the outlet size imposes a characteristic length scale in the system. This scaling extends to quantities never explored before with the precision reached here, such as the particle orientation and the order parameter.      

\subsection{Comparison of free-flow with the flow-rate-controlled case}
\subsubsection{Vertical velocity and packing fraction}
\begin{figure*}[!htbp]
    \centering
    \includegraphics[width=1.0\textwidth]{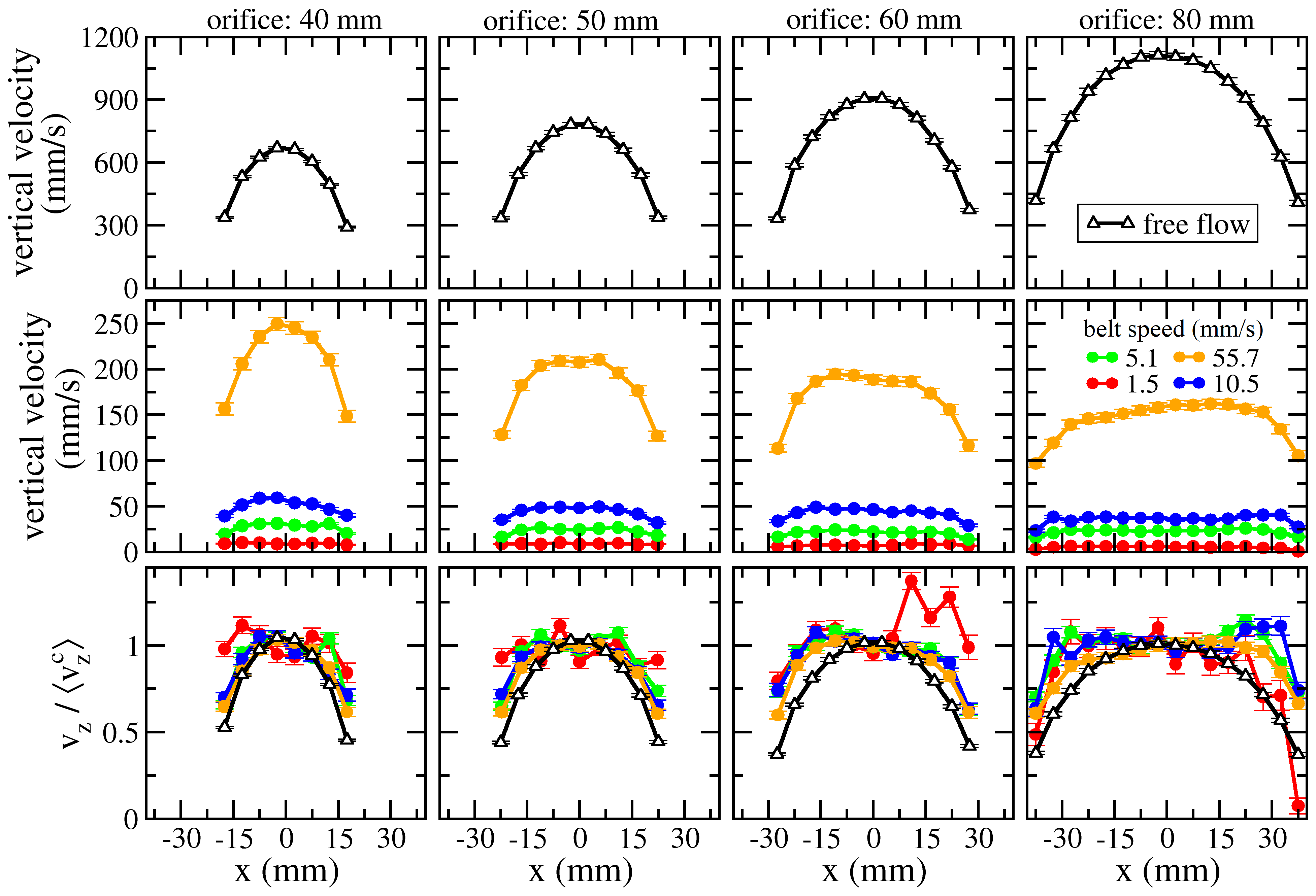}
    \caption{Vertical velocity at the orifice for different orifice sizes (from left to right: $40$ mm, $50$ mm, $60$ mm and $80$ mm). Top row: black lines with open triangular symbols represent the case of free flow. Middle row: colored lines with solid circular symbols represent the cases of flow-rate-controlled flow with different conveyor belt speeds ($1.5$ mm/s, $5.1$ mm/s, $10.5$ mm/s, $55.7$ mm/s) as indicated in the legend of the top right panel. Bottom row: vertical velocity normalized by the central mean vertical velocity $\langle v^c_z\rangle$.}
    \label{fig:v_z}
\end{figure*}

\begin{figure*}[!htbp]
    \centering
    \includegraphics[width=1.0\textwidth]{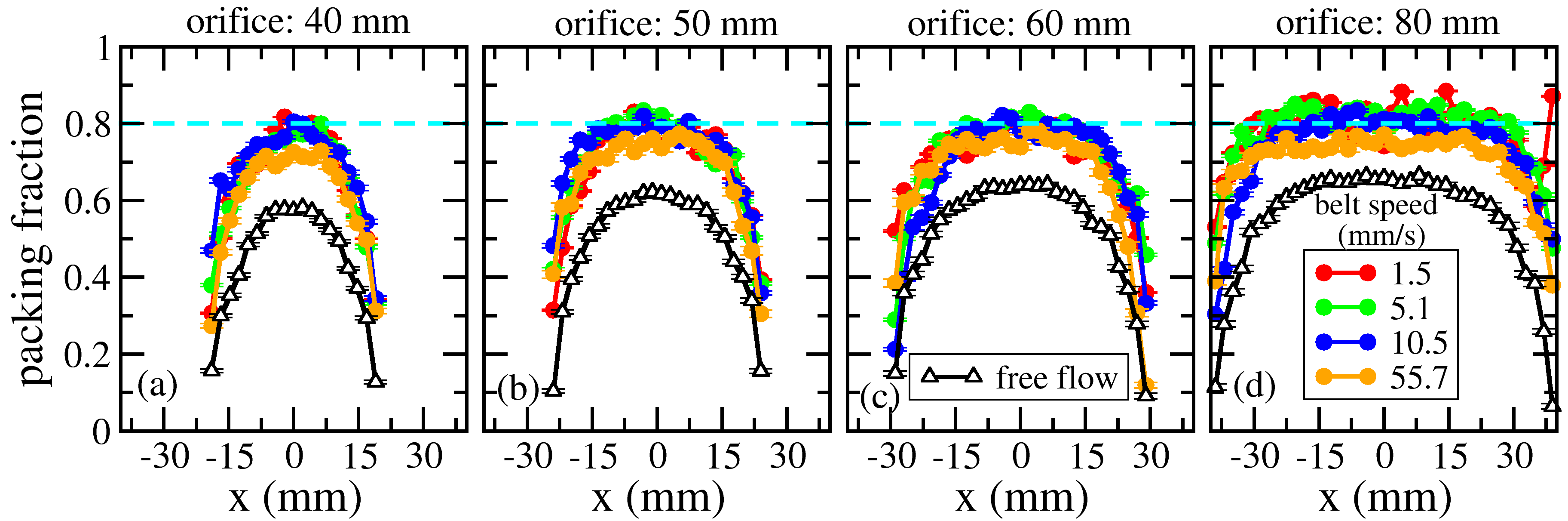}
    \caption{Packing fraction at the orifice for different orifice sizes (from left to right: $40$ mm, $50$ mm, $60$ mm and $80$ mm). The dashed line is a guide at the packing fraction of $0.8$. Black lines with open triangular symbols represent the case of free flow. Colored lines with solid circular symbols represent the cases of flow-rate-controlled flow with different conveyor belt speeds ($1.5$ mm/s, $5.1$ mm/s, $10.5$ mm/s, $55.7$ mm/s) as indicated in the legend of panel (d).}
    \label{fig:packing_fraction}
\end{figure*}
After the analysis of the free discharges we now compare these results with the conveyor belt discharges. First, we analyze the vertical velocity at the outlet line. The profiles for different orifice sizes (from left to right: $40$ mm, $50$ mm, $60$ mm and $80$ mm) are plotted in Fig.~\ref{fig:v_z} where we show results for the free-flow case (top row), and the flow-rate-controlled cases with different belt speeds, both in physical units (middle row) and in normalized form (bottom row). 
\begin{figure}[htbp!]
    \centering
    \hspace*{-1cm}\includegraphics[width=0.9\linewidth]{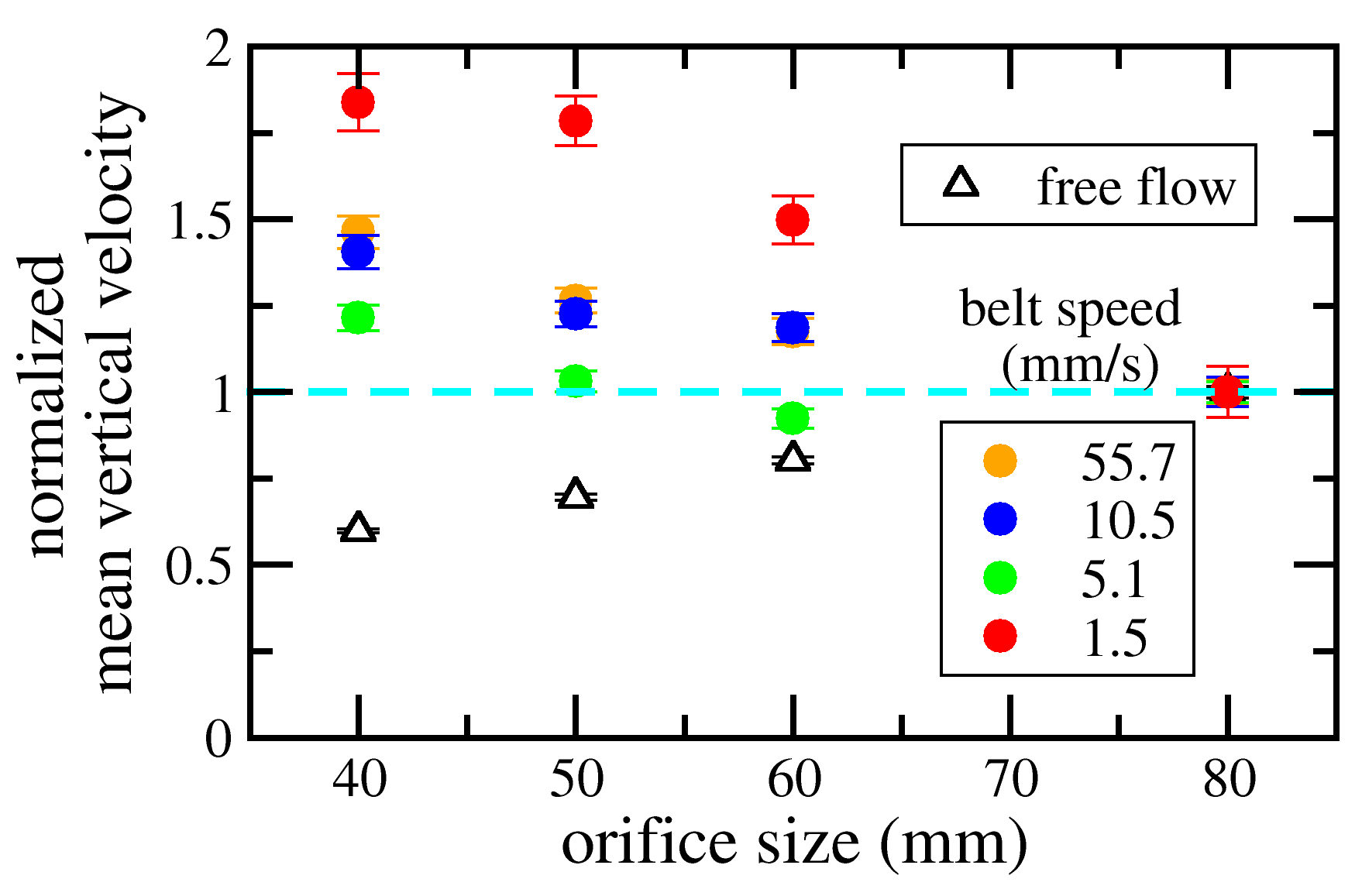}
    \caption{Normalized mean vertical velocity as a function of orifice size. The average vertical velocity values are normalized with the corresponding value for the orifice of $80$ mm. Open triangles represent the case of free flow. Colored solid circles represent the flow-rate-controlled cases as indicated in the legend. The dashed line is a guide at the normalized value of $1$.}
    \label{fig:mean_vz}
\end{figure}

As expected, for all orifices the vertical velocity rises as the belt speed grows (see middle row). 
More counterintuitively, we observe that for the highest belt speed (orange curves), the mean vertical velocity at the outlet reduces as the orifice size enlarges. This behavior also occurs for the other belt speeds as it becomes evident in Fig.~\ref{fig:mean_vz} where we represented, for each belt speed, the mean vertical velocity normalized by the value it takes for the largest orifice (80 mm). Remarkably, the reduction of the vertical velocity with the outlet size contrasts the increasing trend observed for the free-flow case, but is in good agreement with the results reported for spheres \cite{gellaPT2020}. The reason given in that work is based on mass conservation and the fact that flow expands after crossing the orifice line, i.e. the belt removes a wider band of grains than the orifice size, and the relative strength  of this effect is larger for a smaller orifice. The reason for the expansion below the orifice is twofold: (i) the trapezoid shape of the region below the orifice line and (ii) the gap between the bottom of the silo wall and the belt (see the enlarged orifice region at the bottom of Fig.~\ref{fig:setup}(a)).

\begin{figure*}[htbp!]
    \centering
    \includegraphics[width=1.0\textwidth]{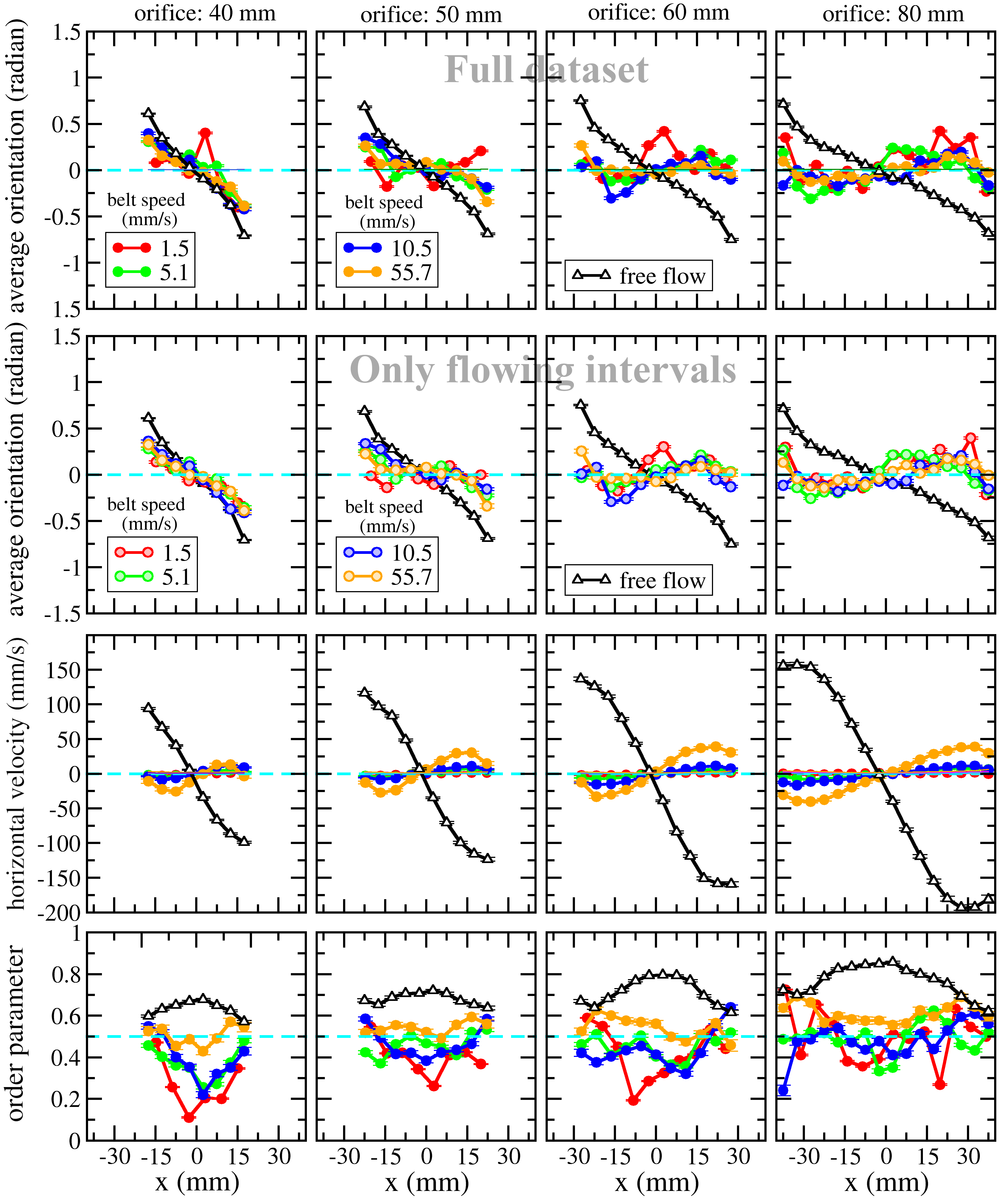}
    \caption{Top two rows: average particle orientation at the orifice, with all data included (first row) and data only from flowing time intervals (second row). The orientation is defined as the angle in radian between the vertical line and the particle's major axis. Third row: horizontal velocity at the orifice. Black lines with open triangular symbols represent the case of free flow. Colored lines with solid circular symbols represent the cases of flow-rate-controlled flow with different conveyor belt speeds ($1.5$ mm/s, $5.1$ mm/s, $10.5$ mm/s, $55.7$ mm/s) as indicated in the legend. Bottom row: Orientational order parameter at the orifice. The dashed line is a guide at $0.5$ that can be taken as a reference to visualize the profiles evolution as the orifice size increases.}
    \label{fig:orientation_order}
\end{figure*}

By normalizing the vertical velocity with the maximum value (developed in the middle region of the orifice) we get the curves presented in the bottom row of Fig. \ref{fig:v_z}. As it could be envisaged from the profiles reported in the medium row, the collapse is not good at all because the curves flatten as the orifice enlarges. We also note that the normalized vertical velocity profiles become increasingly noisy with decreasing belt speed. This behavior is attributed to the augment of the relative velocity fluctuations when reducing belt speed described before.

The packing fraction curves at the orifice line for free flow and the belt extraction system are compared in Fig. \ref{fig:packing_fraction}. The first thing to note is that decreasing belt speed leads to higher packing fraction as it was already reported for spheres \cite{gellaPT2020}. In the same way, we find increasing packing fractions with enlarging the orifice sizes (see the relative position of the colored curves to the dashed cyan guide line). This is in accordance with the above described observation of decreasing flow velocity with increasing orifice size (Fig. \ref{fig:v_z}, middle row). Although the noisy nature of the velocity curves (see bottom row of Fig. \ref{fig:v_z}) does not allow a meaningful calculation of the shear rate, the increasing trend of the packing fractions with decreasing belt velocity suggests that packing fraction increases with decreasing shear rate, just like for the free flow case (Fig. \ref{shear_rate-phi-s}(a)).
At this stage it should be recalled that, in the belt discharges, the flow becomes increasingly intermittent with decreasing belt speed. Therefore, as the packing fraction curves contain data from both flowing and arrested configurations, the smaller the belt velocity, the higher the number of times in which the flow is arrested that are used to compute the time averaged packing fraction profiles. Accordingly, the average packing fraction values obtained increase as the belt velocity reduces. 

\subsubsection{Orientation and order parameter}

\begin{figure}[htbp!]
    \centering
    \includegraphics[width=1.0\linewidth]{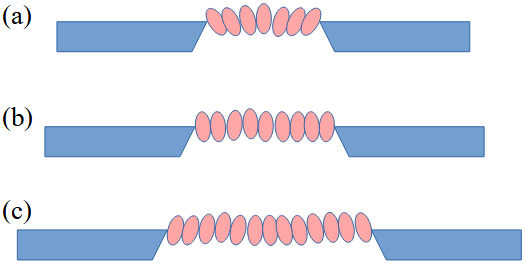}
    \caption{Schematic illustration of the average orientation of the particles at the orifice for the flow-rate-controlled case with increasing the orifice size from (a) to (c).}
    \label{fig:dgm_orientation}
\end{figure}

Next, in Fig.~\ref{fig:orientation_order} we present the profiles at the outlet of i) the average orientation, ii) the horizontal velocity and iii) the order parameter. As for the other quantities, we show results for different orifice sizes and for both, free-flow and several flow-rate-controlled scenarios. 
For the average orientation we present two versions. The first one includes all data (top row) while the second one includes data only from the flowing time intervals (thus arrested time intervals excluded). 
Both versions show the same general trend: when the orifice is relatively small (40 mm) we observe similar behavior in the belt discharged system as in the free flow case: the average orientation decreases more or less linearly as $x$ increases, taking positive values when $x<0$ and negative values when $x>0$. This implies that in average grains point towards the center of the orifice as shown in Fig.~\ref{fig:dgm_orientation}(a). However, as the orifice enlarges, this trend is altered and the differences among the belt discharges and the free case become increasingly larger. The most important difference is observed for $D=80$ mm, where for the belt controlled case, the average orientation of the grains passing through the orifice is rather vertical, or even slightly pointing towards the two edges as shown in Fig.~\ref{fig:dgm_orientation}(c). This could be attributed to the expansion of the flow below the orifice explained above. The average grain alignment is sketched in the presence of the belt for small, middle and large orifices in Figs.~\ref{fig:dgm_orientation}(a), (b) and (c), respectively. In line with this, the profiles of the horizontal velocity (Fig.~\ref{fig:orientation_order}, third row) reveal that while for the free flow case the grains at the orifice line move towards the center (velocity is positive when $x<0$ and negative when $x>0$), for the belt discharges they move towards the edges (velocity is negative when $x<0$ and positive when $x>0$). Again, this result corroborates the diverging nature of the flow below the orifice line when the silo is discharged by the belt.

Additionally, the curves of the average particle orientation (Fig.~\ref{fig:orientation_order}, top two rows) reveal that the flow-rate-controlled cases are not as smooth as the ones in the free-flow case, and their smoothness gets improved once the belt speed reaches the highest value (orange curves). The higher noise on the orientation profile for lower belt speeds is related to the intermittent nature of the flow. This is evidenced by the fact, that the curves in the second row, where the datasets originate from the flowing time intervals only (thus arrested time intervals excluded), are smoother compared to the curves in the the top row where all data are included. This is especially true for the lowest belt speed. This observation is consistent with Fig.~\ref{fig:flow_rate-orifice-belt}(d) and the panels of the last column in Fig.~\ref{fig:flow_field_maps}, where we see that the relative fluctuations in flow rate and velocity become significantly larger when the belt speed decreases to $1.5$ mm/s. Consequently, when the flow stops, grain orientations are perturbed. 

Regarding the order parameter (bottom row in Fig.~\ref{fig:orientation_order}), 
we globally find lower values for the case of belt controlled flows than for the free flow, indicating that the presence of the belt destroys the orientational order. Intriguingly, the curves of the free-flow case have a convex shape for all the orifice sizes, with the largest order parameter in the middle of the orifice, while the ones for the flow-rate-controlled cases $-$ especially for low belt speed $-$ have a dip in the middle. We attribute this feature to the following behavior. In the middle of the orifice the average orientation of the particles is vertical for both free flow and belt controlled flow. However, for the case with belt, when the system is arrested due to the intrinsic intermittencies of the flow, the grains tend to turn (left or right with equal probability). As a consequence, the order parameter reduces while the average orientation remains vertical. In this sense, the concavity of the order parameter curves can be seen as an indicator of the degree of flow intermittencies developed within the system.   

\section{Summary}
Our results show that the presence of a conveyor belt strongly modifies the flow dynamics of elliptical grains inside the silo. We find that limiting the flow with the belt leads to a larger stagnant zone in the two sides of the silo. Decreasing the belt speed naturally leads to decreasing flow velocity in the silo, but the amplitude of velocity fluctuations decreases much more slowly than that of the average velocity. This means that the relative velocity fluctuations defined as the ratio of velocity fluctuations and average flow velocity are considerably increased in most parts of the silo when we are limiting the discharge with the belt. In other words, decreasing belt speed leads to fluctuating, intermittent flow instead of smooth but slower flow. This effect (already known for spheres) leads to \textit{a priori} unexpected behavior of the order parameter, which can be seen as an indicator of the degree of intermittencies developed in the system. 

Another interesting feature introduced by the presence of the belt and the geometry of our system is the lateral spreading of the grains below the orifice. This behavior was also reported for spherical particles and can be clearly detected looking at the profiles of horizontal velocities at the outlet: for free discharges, grains move towards the center of the orifice, while for discharges with the belt, grains move towards the edges. Interestingly, for the case of particles with anisotropic shape, the lateral spreading of the grains has a side effect that affects their average orientation: while for free discharge the alignment was pointing towards the center of the orifice, for belt limited discharge decreasing belt speed and increasing orifice size leads to a change in the average alignment, which becomes vertical or even slightly inclined towards the orifice edges.

\section*{Acknowledgements}
We thank Prof. Diego Maza, Diego Gella and Luis Fernando Urrea for their help in the initial stages of this experiment.
The authors acknowledge financial support from the European Union's Horizon 2020 Marie Sk\l{}odowska-Curie grant ''CALIPER'' (No. 812638) and Spanish Government Project PID2020-114839GB-I00 funded by MCIN/AEI/10.13039/501100011033.

\bibliographystyle{apsrev4-1}
\bibliography{references}

\end{document}